\documentclass[apj]{emulateapj}
\usepackage{apjfonts}

\begin{document}

\newcommand{\CIV}{C~{\sc iv}}
\def\sarc{$^{\prime\prime}\!\!.$}
\def\arcsec{$^{\prime\prime}\, $}
\def\arcmin{$^{\prime}$}
\def\kms{${\rm km\, s^{-1}}$}
\def\degr{$^{\circ}$}
\def\seco{$^{\rm s}\!\!.$}
\def\ls{\lower 2pt \hbox{$\;\scriptscriptstyle \buildrel<\over\sim\;$}}
\def\gs{\lower 2pt \hbox{$\;\scriptscriptstyle \buildrel>\over\sim\;$}}

\def\mbh{$M_{\rm BH}$}
\def\mstar{$M_{\rm STAR}$}
\def\sis{$\sigma$}
\def\vvir{$V_{\rm vir}$}
\def\ms{$M_{\rm STAR}$}
\def\rhoz{$\rho_{\bullet}(z)$}
\def\rhovdf{$\rho_{\rm VDF}(z)$}

\def\gsim{\;\rlap{\lower 2.5pt
 \hbox{$\sim$}}\raise 1.5pt\hbox{$>$}\;}
\def\lsim{\;\rlap{\lower 2.5pt
   \hbox{$\sim$}}\raise 1.5pt\hbox{$<$}\;}

\title{The Evolution of the \mbh-\sis$\,$ relation Inferred from the Age Distribution of\\Local Early-Type Galaxies and AGN Evolution}

\author{Francesco Shankar\altaffilmark{1}, Mariangela Bernardi\altaffilmark{2}, and Zolt\'{a}n Haiman\altaffilmark{3}}
\altaffiltext{1}{Department of Astronomy,  The Ohio State University, Columbus, OH 43210; shankar@astronomy.ohio-state.edu}
\altaffiltext{2}{Department of Physics and Astronomy, University of Pennsylvania, 209 South 33rd St, Philadelphia, PA 19104; bernardm@physics.upenn.edu}
\altaffiltext{3}{Department of Astronomy, Columbia University, 550 West 120th Street, New York, NY 10027; zoltan@astro.columbia.edu}


\begin{abstract}
We utilize the local velocity dispersion function (VDF) of
spheroids, together with their inferred age--distributions, to
predict the VDF at higher redshifts ($0<z\lsim 6$), under the
assumption that (i) most of the stars in each nearby spheroid formed
in a single episode, and (ii) the velocity dispersion \sis\ remained
nearly constant afterward. We assume further that a supermassive
black hole (BH) forms concurrently with the stars, and within $\pm$
1 Gyr of the formation of the potential well of the spheroid, and
that the relation between the mass of the BH and host velocity
dispersion maintains the form $M_{\rm BH} \propto \sigma^{\beta}$
with $\beta\approx 4$, but with the normalization allowed to evolve
with redshift as $\propto (1+z)^{\alpha}$. We compute the BH mass
function associated with the VDF at each redshift, and compare the
accumulated total BH mass density with that inferred from the
integrated quasar luminosity function (LF; the so--called So\l tan
argument).  This comparison is insensitive to the assumed duty cycle
or Eddington ratio of quasar activity, and we find that the match
between the two BH mass densities favors a relatively mild redshift
evolution, with $\alpha\sim 0.26$, with a positive evolution as
strong as $\alpha\gtrsim 1.3$ excluded at the 99\% confidence level.
A direct match between the characteristic BH mass in the VDF--based
and quasar LF--based BH mass functions also yields a mean Eddington
ratio of $\lambda\sim 0.5-1$ that is roughly constant within
$0\lesssim z \lesssim 3$. A strong positive evolution in the
\mbh-\sis\ relation is still allowed by the data if galaxies
increase, on average, their velocity dispersions since the moment of
formation, due to dissipative processes.
If we assume that the
mean velocity dispersion of the host galaxies evolves as
$\sigma(z)=\sigma(0)\times(1+z)^{-\gamma}$, we find a lower limit of
$\gamma\gtrsim 0.23$ for $\alpha\gtrsim 1.5$. The latter estimate
represents an interesting constraint for galaxy evolution models and
can be tested through hydro simulations. This dissipative model,
however, also implies a decreasing $\lambda$ at higher $z$, at
variance with several independent studies.
\end{abstract}


\section{Introduction}
\label{sec|intro}

It has now been assessed that most, if not all, local galaxies have a
supermassive black hole (BH) at their center, the mass of which is
tightly correlated with the velocity dispersion \sis$\,$ and other bulk
properties of the host
galaxy (e.g., Ferrarese \& Merritt 2000; Gebhardt et al. 2000).
However, the sample of local galaxies for which the BH sphere of
influence has been resolved amounts to only $\sim 30$.  It is not
clear how representative this small sample is of the whole BH population,
and whether the correlations seen in the sample already held in the past.

Peng et al. (2006) have collected a sample of 31 lensed and 18
non-lensed Active Galactic Nuclei (AGN) at redshifts $z>1.7$. They
measured rest-frame $R$-band luminosities from $H$-band fluxes and
BH masses by applying virial relations based on emission line
widths. They found that the BH-to-host galaxy luminosity at $z\sim
2$ is about the same as the one at $z\sim 0$. Therefore, once the
observed rest-frame luminosity is dimmed through passive evolution
to $z\sim 0$, at fixed BH mass the ratio BH-to-host luminosity grows
significantly,
and the resulting BH-luminosity normalization is several times
higher than the local one. Similar results were derived by McLure et
al. (2006), who measured the BH-to-host galaxy mass ratio in a
sample of radio-loud AGNs in the redshift range $0<z<2$ finding
\mbh/\ms$\propto (1+z)^2$. Shields et al. (2006) found that the CO
emission lines in a sample of $z>3$ quasars is very narrow,
suggesting bulge mass about an order of magnitude lower than
measured in the local universe, at fixed BH mass (see also Coppin et
al. 2008). Treu et al. (2007) found that the BH masses in a sample
of 20 Seyferts galaxies at $z=0.36$ are offset by an amount of
$\Delta \log$\mbh$\sim 0.5$ at fixed velocity dispersion, which
implies an evolution of \mbh/\ms$\propto (1+z)^{1.5\pm 1.0}$,
consistent with that derived by the previous works.

On the other hand, Lauer et al. (2007) have discussed several
possible biases which may seriously affect these findings. At high
redshifts a sample will be biased toward the most luminous AGNs and
more massive BHs. Given the observed scatter in the local relations,
especially significant in the \mbh-host luminosity relation, these
massive BHs will be preferentially associated with the less massive,
but more numerous galaxies, yielding a false sign of evolution. When
the cumulative mass density of AGNs is taken into account, several
authors (e.g., Haiman et al. 2004; Marconi et al. 2004; Silverman et
al. 2007; Shankar et al. 2008a, hereafter SWM) have shown that once
rescaled by a simple constant, it provides a good match to the
cosmological star formation rate density. De Zotti et al. (2006) and
SWM have shown that the galaxy stellar mass function at $z\sim 2$,
mostly composed of massive early-type galaxies (e.g., Drory et al.
2005), converted into a BH mass density assuming a \mbh/\ms$\,$
ratio 3-5 times higher than in the local universe, would imply a BH
mass density already close, if not higher, than that inferred in the
local universe, leaving no room for further accretion at $z\lesssim
2$, where, in fact, a significant fraction of the total AGN energy
output is produced. Recently, Ho et al. (2008) compiled a sample of
154 nearby ($z<0.1$) active galaxies showing substantial ongoing BH
growth in the most actively accreting AGNs, where BH growth appears
to be delayed with respect to the assembly of the host galaxy.

In this paper, we propose a simple, yet robust, way to constrain the
degree of redshift evolution in the \mbh-\sis$\,$ relation, that is
relatively insensitive to assumptions that relate the SMBH
population to quasars. We combine the measured velocity dispersion
function (VDF) of local spheroids with a postulated power--law
redshift--dependence of the \mbh-\sis$\,$ relation. By comparing the
resulting total BH mass density at each redshift with the same
quantity inferred from integration of the active galactic nuclei
(AGN) luminosity function (see So\l tan 1982), we find the degree of
evolution required in the \mbh-\sis$\,$ relation to match these two
independent estimates. This approach yields results based on the
``bulk'' of the active BHs at all redshifts, and is therefore
relatively insensitive to possible biases which may affect studies
performed on small samples of high--redshift luminous quasars (e.g.,
Lauer et al. 2007). After describing the sample used in our
computations in \S~\ref{sec|data}, we proceed to derive our main
results in \S~\ref{sec|results}. These results are discussed further
in \S~\ref{sec|conclu}, where we also offer our conclusions.
Throughout this paper we use the cosmological parameters
$\Omega_m=0.30$, $\Omega_\Lambda=0.70$, and $h\equiv H_0/100\, {\rm
km\, s^{-1}\, Mpc^{-1}}=0.7$, consistent with the three-- (Spergel
et al. 2007) and five--year (Dunkley et al. 2008) data from the {\it
Wilkinson Microwave Anisotropy Probe (WMAP)}.

\section{DATA}
\label{sec|data}

We have used the sample of early-type galaxies obtained by Bernardi
et al. (2006). The sample, extracted from the Sloan Digital Sky
Survey (York et al. 2000), contains over 40,000 early-type galaxies,
selected for having an apparent magnitude $14.5\lesssim M_r \lesssim
17.75$, extending over a redshift range $0.013 < z < 0.25$, which
corresponds to a maximum lookback time of 3 Gyr.  The ages of
galaxies are computed in two different ways, discussed in detail by
Jimenez et al. (2007), from (i) single stellar population spectral
fitting, using the MOPED algorithm (Heavens et al. 2000) to
determine the full star--formation history of the galaxies, and (ii)
using the published ages by Bernardi et al. (2006) which were
obtained by fitting the Thomas et al. (2005) $\alpha$-enhanced
models to the Lick index absorption features measured from stacked
spectra of galaxies with similar properties. The age--distributions
at fixed velocity dispersion $\sigma$ are generally broad, but tend
to be narrower and centered on older ages for higher values of
$\sigma$.  Such effects are more marked for age distributions
inferred from MOPED (see Figure 1 in Haiman et al. 2007).  We will
compare results obtained by adopting either the MOPED or the
Lick-index age distribution in \S~\ref{sec|results}.

\begin{figure*}[th]
\includegraphics[angle=0,scale=2.]{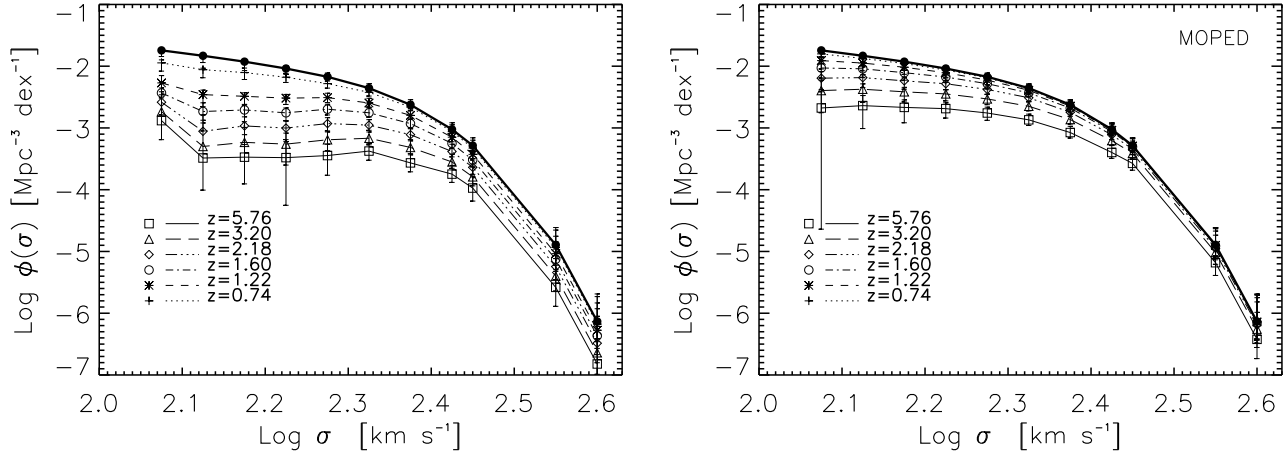}
\centering \caption{Velocity dispersion function (VDF) at
  different redshifts, as labeled, obtained by combining the local VDF
  with the age distribution of local galaxies
  (equation~[\ref{eq|VDFz}]). \emph{Left panel}: VDF obtained using
  the ages computed from Lick indexes; \emph{Right panel}: VDF
  obtained using the ages computed from the MOPED algorithm (see text
  for details).} \label{fig|VDF}
\end{figure*}

The analysis presented in Bernardi et al. (2006) and Haiman et al.
(2007) probe velocity dispersions within $2.05\lesssim \log(\sigma/{\rm km\, s^{-1}}) \lesssim 2.45$.
Here we extend such
analysis including the age distributions of galaxies with velocity
dispersion $2.45\lesssim \log(\sigma/{\rm km\, s^{-1}})\lesssim
2.55$. We find that galaxies within this last bin are even older
than the oldest galaxies probed by Haiman et al. (2007), confirming
and extending the general trend of increasing age for larger
$\sigma$. Instead of considering one single bin with mean velocity
dispersion $\log(\sigma/{\rm km\, s^{-1}})=2.5$, we have treated the
bin as two distinct bins with $\log(\sigma/{\rm km\, s^{-1}})=2.45$
and $2.55$ which we have assumed share the same age distributions as
the total bin. We have also included an additional bin with
$\log(\sigma/{\rm km\, s^{-1}})=2.60$, which we have again assumed to have
an age distribution equal\footnote{a more appropriate choice would
be to assign older ages to the galaxies with extreme velocity
dispersions, given the general trend of older ages for higher
$\sigma$, however this would pose even stronger evidence for the
downsizing discussed below further strengthening our general
conclusions.} to the one with $\log(\sigma/{\rm km\, s^{-1}})=2.45$
and $2.55$ (a direct estimate of the ages for these galaxies with
the techniques discussed above is highly limited by the low
Signal-to-Noise of the spectra). As will be shown in
\S~\ref{sec|results} (Figure~\ref{fig|VDF}), this binning in $\log
\sigma$ enables us to better probe the statistical evolution of the
VDF even at large velocity dispersions, and it has a negligible
effect in the resulting cumulative black hole mass density and on
our general results.

\section{RESULTS}
\label{sec|results}

We first estimate the VDF as a function of redshift $z$. At any $z$,
the VDF in a given bin of velocity dispersion $\sigma_i$ is given by
all the galaxies which have formed prior to $z$. Therefore to
compute $\Phi(\sigma_j,z)$ we subtract from the local census of
galaxies with velocity dispersion $\sigma_i$ those galaxies that
have an age $\tau$ lower then the lookback time $\tau_j(z)$,
\begin{equation}
\Phi(\sigma_i,z)=\left[1-\sum_{\tau<\tau_j}p(\tau_j(z)|\sigma_i)\right]\times
\Phi(\sigma_i)\, . \label{eq|VDFz}
\end{equation}
Note that $p(\tau_j(z)|\sigma_i)$ refers to the fraction of galaxies
with velocity dispersion $\sigma_i$ which have an age of
$\tau_j(z)\pm$ 1 Gyr. Therefore $\Phi(\sigma_i,z)$ includes in the
$\sigma_i$ bin all galaxies whose ages are within $\pm 1$ Gyr of
$\tau_j(z)$. The VDF at $z=0$ is taken from Sheth et al. (2003) and
includes the contribution of bulges of spirals. We therefore assume
that bulges of spirals and local spheroids within the same bin of
velocity dispersion share similar age distributions. However, as
discussed in \S~\ref{sec|conclu} below, our results would still hold
even if the contribution from spirals were neglected.  The
statistical uncertainties associated with $\Phi(\sigma,z)$ are
computed from equation~(\ref{eq|VDFz}) through error propagation
including uncertainties in $p(\tau_j(z)|\sigma)$, given by Haiman et
al. (2007; see their Figure 1) and $\Phi(\sigma)$, given by Sheth et
al. (2003).

Figure~\ref{fig|VDF} shows the VDF obtained from
equation~(\ref{eq|VDFz}).  The different curves in both panels show
$\Phi(\sigma,z)$ at different redshifts, as labeled.  At fixed
redshift, the symbols indicate the position of the mean in the bin
of $\log \sigma$ considered, for which reliable age distributions
$p(\tau_j(z)|\sigma_i)$ have been computed. In the left panel of
Figure~\ref{fig|VDF}, the $p_{ji}$ distributions have been derived
from the Lick--indices method, while the right panel shows the
results with the MOPED-based $p_{ji}$ distributions. In our analysis
below, we will adopt the $\Phi(\sigma,z)$ implied by the $p_{ji}$
derived from Lick indices. However, we will discuss the consequences
of the alternate choice on our results in \S~\ref{sec|conclu}. In
both cases, we reproduce the conclusion of previous work (e.g.,
Trager et al. 2000; Thomas et al. 2005; Bernardi et al. 2006;
Jimenez et al. 2007; Haiman et al. 2007) -- that is, we find strong
evidence for \emph{downsizing}: on average, galaxies with larger
velocity dispersion are formed earlier.  This behavior is expected
from basic galaxy formation theory: high--redshift galaxies form in
a denser universe and therefore preferentially form out of baryonic
clumps collapsed in denser, gas rich environments which in turn,
induce more dissipation, more compact remnants and higher velocity
dispersions. At fixed velocity dispersion, the MOPED ages are higher
than inferred from Lick indices, producing a less pronounced
evolution in the VDF at $0\lesssim z \lesssim 3$. Theoretical models
in which the galaxy velocity dispersion is linked with the virial
velocity of the host halo (e.g., Ferrarese 2002) predict similar
trends for the VDF as a function of time (Cirasuolo et al. 2005; see
also Loeb \& Peebles 2003).

\begin{figure*}[th]
\epsscale{1.} \centering
\plotone{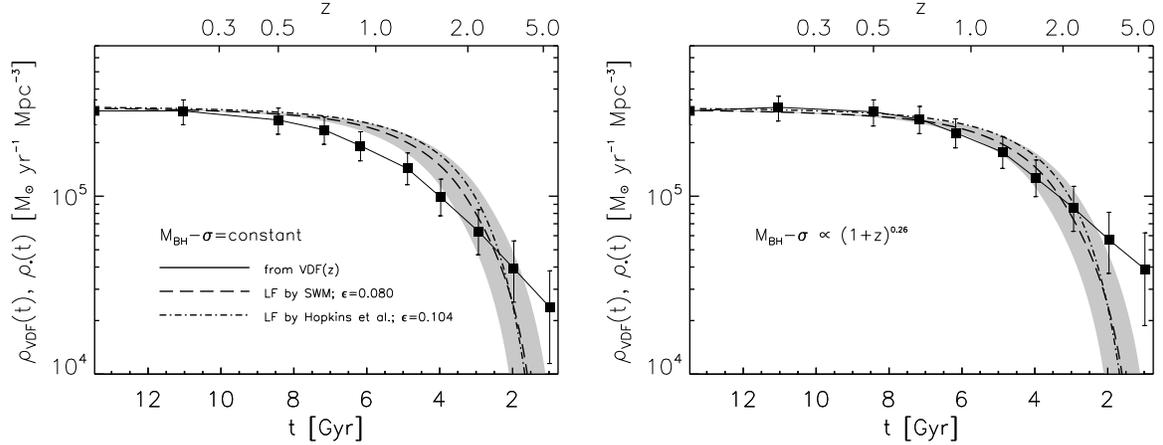} \caption{Comparison between the accreted mass
  density at each redshift obtained from $\Phi(\sigma,z)$, convolved
  with the local \mbh-\sis$\,$ relation (\emph{solid} curve and
  \emph{solid} squares) and the mass density inferred from integration
  of the Shankar et al. (2008a) AGN luminosity function (\emph{long
    dashed} curve) and a radiative efficiency of $\epsilon=0.080$; the
  grey area represents the uncertainty at each time $t$ associated to
  the mass accreted within $t \pm 1$ Gyr; the \emph{dot-dashed} line
  is the predicted accreted mass using the Hopkins et al. (2007) AGN
  luminosity function and a radiative efficiency of
  $\epsilon=0.104$. The \emph{left panel} shows
\rhovdf\ predicted from the VDF assuming that the
  \mbh-\sis$\,$ relation is independent of redshift, while the
  \emph{right panel} shows the predictions for the best-fit
\rhovdf\ when the normalization of the \mbh-\sis\
  relation evolves as $\propto(1+z)^{0.26}$.}
\label{fig|rhoBHz}
\end{figure*}


The BH mass function implied by the VDF at any time is given by
converting $\Phi(\sigma,z)$ to a BH mass function through the
\mbh-\sis$\,$ relation and a convolution with a Gaussian with
intrinsic scatter of $0.22$ dex.
We assume here that $\log$\mbh\ at fixed $\log$\sis\ is given by a
Gaussian distribution, with a mean of
\begin{equation}
\log \left(\frac{M_{\rm BH}}{M_{\odot}}\right)=8.21+3.83\log
\left(\frac{\sigma}{200\, {\rm km\, s^{-1}}}\right)+\alpha
\log(1+z)\, ,
    \label{eq|Mbhz}
\end{equation}
and a standard deviation of $\eta=0.22$.  This latter value represents
the intrinsic scatter as given by Tundo et al. (2007) and as recently
confirmed by Shankar \& Ferrarese (2008). By integrating the resulting
BH mass function at all times, we derive the total BH mass density
\rhovdf, corresponding to BHs in the range of \sis\ and BH mass probed
by our sample at each redshift.

%
We then compare \rhovdf$\,$ with the BH mass density obtained by
direct integration of the AGN luminosity function $\Phi(L,z)$ from
$z=6$ up to redshift $z$. The latter quantity is given by
\begin{equation}
\rho_{\bullet}(>\log L_{\rm min},z)=\frac{1-\epsilon}{\epsilon
c^2}\int_{z}^6 dz'\int_{\log L_{\rm
min}}^{\infty}\Phi(L,z')L\frac{dt}{dz'}d\log L\, .
    \label{eq|soltan}
\end{equation}
Here $\epsilon$ represents the radiative efficiency, and for our
numerical calculation, we adopt the bolometric AGN luminosity
function $\Phi(L,z)$ from SWM (using the LF from Hopkins et al. 2007
gives similar results as discussed below).
At each redshift, we integrate equation~(\ref{eq|soltan}) above the
minimum luminosity, corresponding to the minimum BH mass probed by
the velocity dispersion distribution in our sample via
equation~(\ref{eq|Mbhz}), which is $M_{\rm BH, min}\sim 10^7\,
M_{\odot}$, in the reference model.
The minimum luminosity is then taken to be the Eddington (1922) luminosity corresponding to this BH mass.
As we discuss in \S~\ref{sec|conclu} below, the exact choice of the
lower limit in the integral of equation~(\ref{eq|soltan}) does not
alter our conclusions.

The growth rate of an active black hole of mass \mbh$\,$ is then
$\dot{M}_{\rm BH}=$\mbh$/t_{\rm ef}$, where the \emph{e}-folding
time is (Salpeter 1964)
\begin{equation}
t_{\rm ef}=4\times
10^7\left[\frac{\epsilon(1-\epsilon)^{-1}}{0.1}\right]\lambda^{-1}\, {\rm yr},
    \label{eq|tefold}
\end{equation}
%
where $\lambda$ is the ratio of the luminosity $\epsilon \dot M_{\rm
BH} c^2$ to the Eddington luminosity. Figure~\ref{fig|rhoBHz}
compares the two independent estimates of BH mass densities. The
accreted mass density at each redshift obtained from
$\Phi(\sigma,z)$ and the \mbh-\sis$\,$ relation is shown with a
solid curve.  The solid squares show the redshifts where the mass
density was computed. The long-dashed curve represents the mass
density inferred from integration of the SWM AGN bolometric
luminosity function. Given that the ages of galaxies in the sample
have a median associated uncertainty of $\pm$ 1 Gyr, at any time
$t(z)$ the BH mass density from AGNs to be compared to \rhovdf$\,$
is systematically uncertain by the mass accreted within $t \pm 1$
Gyr, which we show as the gray area.\footnote{Note that the $t \pm
1$ Gyr uncertainty is for \rhovdf. However, in our calculations,
assigning the uncertainty to $\rho_{\bullet}(z)$ or \rhovdf$\,$
makes no difference. If the time of formation of the galaxies is
uncertain by $\pm 1$ Gyr, then statistically the \rhovdf at the time
$t$ can be compared with the cumulative mass accreted at any time
$t\pm$ 1 Gyr.}
We choose a constant mean radiative efficiency of $\epsilon=0.080$,
which provides a good match to the BH mass density at $z=0$
(e.g. Haiman et al. 2004, SWM). It can be immediately inferred from
the left panel, which assumes an unevolving \mbh-\sis\ relation, that
\rhovdf$\,$ and \rhoz$\,$ are consistent with each other within
errors, and therefore a strong evolution with redshift in the
\mbh-\sis\ relation is not required. Very similar results are found if
we adopt the bolometric luminosity function from Hopkins et al.(2007),
shown as the dot-dashed curve in the same Figure. In this case, we use
a slightly higher radiative efficiency of $\epsilon=0.104$ to
renormalize the total $z=0$ accreted mass density to the local value,
due to the fact that the bolometric corrections used by Hopkins et
al. (2007) are about $30\%$ higher then those adopted by
SWM. Nevertheless, even in this case we find that \rhoz$\,$ well
matches \rhovdf$\,$ at all times.

\begin{figure}[t]
\epsscale{1.} \centering
\plotone{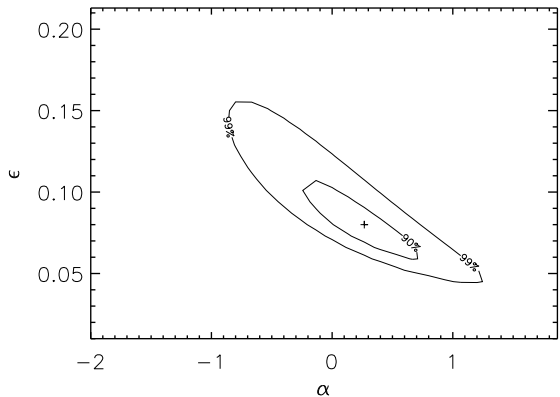} \caption{Confidence levels of 90\% and 99\%
  computed assuming $\chi^2=\chi^2_{\rm min}+2.30$ and
  $\chi^2=\chi^2_{\rm min}+9.21$, respectively, for two parameters in
  the model, the radiative efficiency $\epsilon$ and the exponent
  $\alpha$, where the normalization of the \mbh-\sis\ relation evolves
  as $\propto(1+z)^{\alpha}$. The cross marks the best--fit values of
  $\epsilon=0.08$ and $\alpha=0.26$ corresponding to the minimum
  $\chi^2_{\rm min}\sim 2.6$ (for 8 degrees of freedom). It is evident
  that a strong redshift evolution in the \mbh-\sis$\,$ relation is
  ruled out at a high confidence level, if the radiative efficiency is
  constant in time.} \label{fig|Chi2}
\end{figure}

\begin{figure}[th]
\epsscale{1.} \centering
\plotone{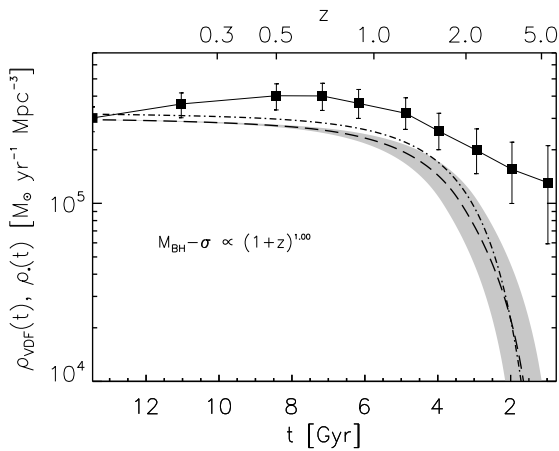} \caption{Same format as Figure~\ref{fig|rhoBHz},
with parameters $\alpha=1$ and $\epsilon\sim 0.08, 0.10$ for the SWM
and Hopkins et al. (2007) luminosity functions, respectively.}
\label{fig|rhoBHz2}
\end{figure}

Joint confidence levels on the two parameters $\epsilon$ and $\alpha$,
inferred from a $\chi^2$ analysis are shown in
Figure~\ref{fig|Chi2}. The cross marks the best--fit model with
$\epsilon=0.08$ and $\alpha=0.26$ (corresponding to the minimum
$\chi^2_{\rm min}\sim 2.6$ for 8 degrees of freedom), which is shown
in the right panel of Figure~\ref{fig|rhoBHz}. Once a constant
radiative efficiency is fixed to match the $z=0$ local and accreted
mass densities, it is evident that the available data favor a
relatively mild redshift evolution of the \mbh-\sis$\,$ relation with
$\alpha\lesssim 0.3$, while a strong evolution with $\alpha\gtrsim
1.3$ is ruled at 99\% confidence level. Likewise, negative evolution
with $\alpha\lsim -1$ is ruled out for any choice of $\epsilon$.  We
note that values of $\alpha\gtrsim 0.8$, in fact, yield the unphysical
result that the absolute total BH mass density \emph{increases} from
$z=0$ to $z\gtrsim 0.7$, as shown in Figure~\ref{fig|rhoBHz2}.  The
confidence contours shown in Figure~\ref{fig|Chi2} may therefore
overestimate the maximum allowed value of $\alpha$.

It is clear from equation~(\ref{eq|soltan}) that the accreted BH
mass density does not depend on the assumed duty cycle or Eddington
ratio distribution $\lambda$(\mbh,$z$), apart from a weak dependence
on the latter through the lower limit of the integration. The
strongest dependencies are on the radiative efficiency and on the
bolometric corrections (see also Figure 9 in SWM).  On the other
hand, the Eddington ratio distribution and its evolution with
redshift, can be constrained by comparing the AGN-based and
VDF$(z)$-based differential BH mass functions (rather than comparing
only the integrated quantities). The $\Phi(\sigma,z)$ convolved with
the \mbh-\sis\ relation (equation~\ref{eq|Mbhz}) in fact, predicts
the shape of the BH mass function for \mbh$\gsim M_{\rm BH, min}$.
On the other hand, as extensively discussed in the literature (see
SWM, and related work by, e.g. Cavaliere et al. 1982; Small \&
Blandford 1992; Salucci et al. 1999; Yu \& Tremaine 2002; Marconi et
al. 2004; Shankar et al. 2004), if a mean Eddington ratio
$\lambda=L/L_{\rm Edd}$ is assumed for the active BHs, then through
a continuity equation and an assumed initial condition, the AGN
luminosity function can be directly mapped into a BH mass function
at all times.  The ``break'' in the predicted BH mass function will
then approximately reflect the break $L^*(z)$ in the observed AGN
luminosity function, i.e., \mbh$^*(z)\propto L^*/\bar\lambda$, where
$\bar\lambda$ is the mean Eddington ratio. Following So{\l}tan
(1982) and Salucci et al. (1999), SWM (see also, e.g., Yu \&
Tremaine 2002) showed that constraints on the mean radiative
efficiency and Eddington ratio of BHs can be gained by comparing the
directly measured and the accreted BH mass functions. However, the
BH mass function has been directly measured only locally, so this
comparison can be performed only at $z=0$, and can not be used to
glean information on the evolution of these two parameters.

The left panel of Figure~\ref{fig|MFz} compares the BH mass function
predicted from the combination of $\Phi(\sigma,z)$ and the mildly
evolving best--fit \mbh-\sis\ relation with a normalization
$\propto(1+z)^{0.26}$ (shown as thick curves, with their uncertainty
shown in gray), and the mass function predicted from the AGN
luminosity function of SWM assuming a mean $\bar\lambda=0.6$ (shown
as thin curves).\footnote{Note that we assumed an initial duty cycle
of 0.5 at $z=6$; however the BH mass function at $z\lesssim 3.5$
becomes independent of this assumption. See SWM for further
details.}

Figure~\ref{fig|MFz} shows that up to $z\lesssim 3$, a
\emph{constant} (non--evolving) mean Eddington ratio of $\bar
\lambda=0.6$ provides a good match between the shapes of the
accreted BH mass function and the one computed from the VDF. At the
low--mass end, the VDF-based BH mass function starts being
incomplete, while at the high mass end, a higher intrinsic scatter
in the \mbh-\sis$\,$ relation and/or a more complicated Eddington
ratio distribution may improve the match. Fully matching the two BH
mass functions is beyond the scope of this paper (see SWM for
further analysis). Our aim here is merely to demonstrate that our
simple approach also provides hints on the mean Eddington ratio and
its redshift evolution. Similar results are found switching to the
Hopkins et al. (2007) luminosity function. The right panel of
Figure~\ref{fig|MFz} shows that a good match between the BH mass
functions is recovered on adopting a constant $\lambda=1.0$.
Although systematic uncertainties in the bolometric AGN luminosity
function preclude tighter constraints on the mean Eddington ratio
(see SWM for further discussions on these issues), it is remarkable
that simple models with $0.5\lesssim \lambda \lesssim 1.0$ constant
with redshift, can provide a reasonable match with the VDF-based BH
mass functions. An independent way to constrain the Eddington ratio
distribution and its evolution with redshift can be derived by
matching the halo clustering implied by the redshift dependent model
BH mass function and the observed AGN clustering (Shankar et al., in
preparation). We have also checked that the same values of $\lambda$
provide a good match even at $z\gtrsim 3$, however the large
uncertainties associated to the VDF at these high redshifts prevent
any firm conclusion.

A somewhat different version of the above exercise was performed by
Haiman et al. (2007).  Under the assumption that the duty cycle of
quasar activity is short, Haiman et al. (2007) matched the
instantaneous quasar LF at each redshift to the LF predicted from
\rhovdf, plus an assumed constant (non--evolving) duty--cycle and
Eddington--ratio distribution.  This approach neglects the BH mass
accreted during the luminous quasar phases (or at least any
corresponding variation of the ``quasar light--curve'' caused by the
growth in BH mass), and places a constraint directly on the relation
between quasar luminosity $L$ and host velocity dispersion $\sigma$.
While the $L-\sigma$ relation is essentially a convolution of the
Eddington ratio distribution with the \mbh-\sis\ relation, this
approach cannot be used to study these two relations separately.
Nevertheless, Haiman et al. (2007) found no evidence for any
evolution in the $L-\sigma$ relation with redshift; their fits to
the quasar LF are consistent with a constant $0.3\lsim \bar\lambda
\lsim 0.5$ combined with a non-evolving \mbh-\sis\ relation.  Since
the \mbh-\sis\ relation is indeed found here, independently, to be
non--evolving, this breaks the degeneracy in the result of Haiman et
al. (2007) and also requires that the evolution in the Eddington
ratio distribution be modest.

\section{DISCUSSIONS}
\label{sec|conclu}

\subsection{VARYING THE MODEL ASSUMPTIONS}
\label{subsec|modelassumptions}

We have studied more complicated scenarios where we also allow for
the scatter and/or the slope of the \mbh-\sis\ relation to increase
with redshift. For example, steadily increasing the slope from 3.83
to, say, 5.5, at fixed scatter of 0.22 dex, still implies
$\alpha\sim 0.30$. The left panel of Figure~\ref{fig|ScatterRhoBH}
shows instead the comparison between the \rhoz\ and the \rhovdf,
assuming the scatter increases with redshift
from $\eta=0.22$ at $z=0$ to $\eta=0.40$ at $z=5.7$, the highest
redshifts probed by our sample. It can be seen that the best-fit
model requires $\alpha=0.15$, even lower than what reported in
Figure~\ref{fig|rhoBHz}. This is expected as these models tend to
increase the BH mass density associated with the VDF at a given
redshift, implying an even milder degree of evolution in the
\mbh-\sis\ normalization. The right panel of
Figure~\ref{fig|ScatterRhoBH} compares the implied mass functions
predicted by the same $\eta(z)$-model and by the AGN luminosity
function. Similarly to the best-fit model discussed in
\S~\ref{sec|results}, a good match can be recovered assuming a
constant $\lambda=0.6$. Moreover, steadily increasing the intrinsic
scatter from 0.22 to 0.4 dex significantly improves the match
between the VDF- and AGN-based BH mass functions at both the high
and low-mass ends.

We have also explored different models for the redshift evolution in
the \mbh-\sis\ relation. For example, a model in which a quadratic
term of the type $\delta\times\log(1+z)^2$ is added in
equation~(\ref{eq|Mbhz}) (see Wyithe 2004 for a similar test applied
to the local \mbh-\sis\ relation), produces a good match between
\rhovdf\ and \rhoz\ if $\alpha\sim \delta \sim 0.1$, with
$\chi^2\sim 2.5$, comparable to our best-fit model. We therefore
conclude that although the choice for the redshift evolution model
in equation~\ref{eq|Mbhz} is not unique, alternative solutions will
still provide similar constraints on the net amount of allowed
evolution.

\subsection{THE SCATTER IN THE $M_{\rm BH}-\sigma$ RELATION}
\label{subsec|mbhsigmaScatter}

At each redshift, our approach assumes that new BHs are formed with
a mass that is tightly imposed by the velocity dispersion of the
host galaxy. Increasing the normalization and/or scatter of the
\mbh-\sis\ relation at high redshift therefore induces in the local
universe a finite spread in BH mass at fixed velocity dispersion.
If the evolution is steep, this can exceed the observed scatter
$\eta \lesssim 0.22$ dex. Each panel in Figure~\ref{fig|ScatterBins}
plots as solid lines the median BH mass distribution of 100 Monte
Carlo simulations
corresponding to a given bin of velocity dispersion, as labeled. The
BH masses are derived from the redshift-dependent \mbh-\sis\
relation where the redshifts are randomly extracted from the age
distribution $p_{ij}(\sigma)$ competing to each velocity dispersion.
The long-dashed lines refer to the Gaussian distribution with
$\eta=0.22$ measured by Tundo et al. (2007). Both our models with
constant (upper panels) or evolving (lower panels) scatter still
produce at $z=0$ BH mass distributions at fixed velocity dispersion
comparable to what is observed.
The small off-set in the Gaussian
distributions predicted by our simulations with respect to those
observed is induced by the sampling of higher redshift, more massive
BHs. Models characterized by stronger redshift evolution with
$\alpha \gtrsim 0.3$ will then evidently predict a scatter in the
local \mbh-\sis\ relation much larger than what is actually
observed. Mergers are then required to be a significant component in the
evolution of the BH population in these models, as the Monte Carlo
simulations performed by Peng (2007) show that random BH mergers
will tighten the relations between BH and host galaxy masses at late
times. However, frequent mergers may, on the other hand, predict too
many massive BHs with respect to those seen in the local universe
(see Figure 13 in SWM).

\subsection{SYSTEMATIC UNCERTAINTIES IN THE METHOD}
\label{subsec|systematic}

The main result of this paper is shown in the right panel of
Figure~\ref{fig|rhoBHz}, which demonstrates that a good match
between \rhovdf$\,$ and \rhoz$\,$ can be achieved by assuming a mild
redshift evolution in the \mbh-\sis$\,$ relation with
$\alpha\lesssim 0.3$. These results are based on the age
distributions $p_{ji}$ derived from Bernardi et al. (2006). However,
MOPED-based galaxy ages are, on average, larger at fixed velocity
dispersion, predicting a flatter dependence $\Phi(\sigma,z)$ as a
function of redshift $z$ (see Figure~\ref{fig|VDF}).  This will
correspondingly flatten \rhovdf$\,$ {\it vs} redshift, and decrease
the best-fit $\alpha$, therefore requiring an even milder redshift
evolution in the \mbh-\sis$\,$ relation. A null evolution in the
\mbh-\sis\ relation is expected in basic AGN feedback models (e.g.,
Silk \& Rees 1998), in which a tight correlation derives by imposing
equilibrium between the energy released by the central BH, and the
gas binding energy, linked to the velocity dispersion.

Bernardi et al. (2007), Graham (2007), and Shankar \& Ferrarese
(2008) have discussed selection biases in the available sample of
BHs that may induce systematic uncertainties in the determination of
the local BH mass function. However, our conclusions are not
affected by these uncertainties, because a change in the local BH
mass density would be absorbed in the radiative efficiency
$\epsilon$ (i.e. $\epsilon$ would be modified, to match \rhovdf$\,$
and \rhoz$\,$ at $z=0$, but $\alpha$ would not change). By the same
token, our results are only weakly dependent on whether or not the
bulges of spirals are included in the estimate of the local BH mass
function (a weak dependence arises only because the addition of the
spiral bulges slightly skews the age--distribution of the total
population to younger ages; this becomes increasingly less important
toward higher redshifts, where a progressively smaller fraction of
the total BH mass density is contributed by the low--$\sigma$
galaxies).

Likewise, uncertainties in redshift--independent bolometric
corrections do not alter our conclusions.  The bolometric correction
adopted in SWM is lower by $\sim$ 30\% with respect to the one used
by Hopkins et al. (2007), but the sole effect of this difference is
to yield a proportionally smaller value of the mean radiative
efficiency to recover the match between \rhovdf$\,$ and \rhoz$\,$ at
$z=0$ (see left panel of Figure~\ref{fig|rhoBHz}). Moreover, the
break luminosity and bright-end slopes of the SWM and Hopkins et al.
(2007) luminosity functions are somewhat different (see Figure 4 in
SWM). Nevertheless, within uncertainties, the resulting BH accretion
histories obtained from the two luminosity functions have a similar
behavior with redshift, both placing the same constraint $\alpha
\lesssim 0.3$ for the evolution in the normalization of the
\mbh-\sis\ relation.

On similar grounds, if we assume that the BHs in our sample radiate
at even lower luminosities than the $L_{\rm min}$ considered in the
integral of equation~(\ref{eq|soltan}), our results do not change.
For example, lowering the minimum luminosity to $\log L_{\rm
min}/{\rm erg\, s^{-1}}=42$, the cumulative emissivity of AGNs
increases by about $\sim 30\%$ at all redshifts yielding a very
similar behavior with time. Therefore, a proportionally higher
radiative efficiency plugged into equation~(\ref{eq|soltan}) keeps
the good match with the \rhovdf\, and specifically we find that the
value of $\alpha=0.16$ yields a $\chi^2\sim 2.4$ for 8 degrees of
freedom.

Our conclusions about the (lack of) evolution in the normalization
of the \mbh-\sis\ relation, in general, are more dependent on
redshift--dependent effects. For example, if the bolometric
correction increased to high $z$ (or, e.g., if obscuration were more
significant at higher redshift), this would again further decrease
our favored mild positive redshift evolution in the \mbh-\sis\
normalization. Likewise, evolution in the mean radiative efficiency
and/or the assumed scatter in the \mbh-\sis\ relation would modify
our results, in the sense that our predicted evolution would be
milder if either increased toward high $z$. In principle, to allow
for a stronger evolution in the \mbh-\sis\ relation the radiative
efficiency must significantly decrease at $z\gtrsim 3$ to boost the
accreted mass density at fixed AGN luminous density. However, we
have checked that $\epsilon$ must then rapidly increase at lower
redshifts in order not to overproduce the local BH mass density.
More quantitatively, if we set $\epsilon\sim 0.05$ at $z\gtrsim 3$,
then it must be that $\epsilon\gtrsim 0.05\times[7/(1+z)]^{0.5}$ at
lower redshifts. Such an evolution in $\epsilon$ is not enough to
allow for a strong variation in the \mbh-\sis\ relation. We found
that \rhovdf\ can match the \rhoz\ implied by the
$\epsilon(z)$-model if $\alpha\sim 0.3$, which is close to our
best-fit model. On other grounds, as recently shown by Shankar et
al. (2008b), a too low radiative efficiency at high redshifts seems
to be disfavored by BH accretion models which simultaneously
reproduce the strong quasar clustering measured at $z=3-4$ in SDSS
by Shen et al. (2007a), the mean Eddington ratio of $\lambda \gtrsim
0.5$, measured by Shen et al. (2007b) for the same quasar sample,
and the high redshift quasar luminosity function (e.g., Richards et
al. 2006; Fontanot et al. 2007; Shankar \& Mathur 2007).

\begin{figure*}[t!]
\epsscale{1.} \centering
\plotone{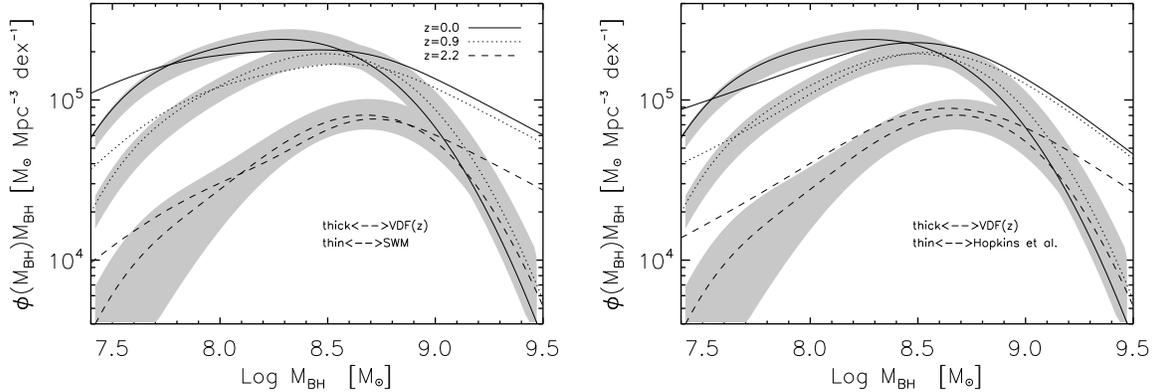} \caption{Comparison between the mass functions
  predicted from $\Phi(\sigma,z)$ convolved with the \mbh-\sis\
  relation whose normalization evolves as $\propto(1+z)^{0.26}$ (thick
  curves, with the uncertainty shown shaded in grey), and the mass
  function predicted from the AGN luminosity functions of Shankar et
  al. (2008; \emph{left panel}) and Hopkins et al. (2007; \emph{right
    panel}). A constant Eddington ratio of
  $\lambda=0.6, 1.0$ has been assumed for computing the accreted mass functions
  in the left and right panels, respectively.
  It can be seen that the choice of a single constant Eddington
  ratio provides a good match to the velocity dispersion-based
  black hole mass functions, at least around the peak of the distributions. A good match is also found
  extending the comparison up to $z\gtrsim 3$, however the large uncertainties at these redshifts prevent any
  firm conclusion.}
 \label{fig|MFz}
\end{figure*}

\subsection{EVOLVING THE MAGORRIAN RELATION}
\label{subsec|magorrian}

Most of the results from other groups discussed in
\S~\ref{sec|intro} focus on the ratio between black hole mass and
stellar \emph{mass}. The latter may settle on longer timescales with
respect to the galaxy velocity dispersion, the amplitude of which is
linked to the central potential well which grows faster than the
overall evolution of the halo (Zhao et al. 2003). In order to get
some hints on the actual evolution of the \mbh-\mstar\ relation with
redshift, we have converted the galaxy stellar mass function into a
BH mass function assuming the \mbh-\mstar\ ratio evolving as
$(1+z)^{\zeta}$. We have used the recent near-infrared stellar mass
function by P\'{e}rez-Gonz\'{a}lez et al. (2008), well constrained
within $0\lesssim z \lesssim 3$ and $10\lesssim \log
M_{\star}/M_{\odot}\lesssim 12$. We have then converted the latter
into a BH mass function by assuming that, on average, about
$0.7\times 10^{-3}(1+z)^{\zeta}$ (e.g., Magorrian et al. 1998;
Marconi \& Hunt 2003) of the total stellar mass is locked up in
spheroids and is associated to the central black hole, with a
Gaussian scatter around the mean of 0.3 dex (e.g., H\"{a}ring \& Rix
2004). In this case, we find that $\zeta \lesssim 0.3$ is a
necessary condition for the BH mass density to be consistent with
the accreted mass from AGNs, the latter derived assuming a fixed
value of the radiative efficiency. This result is in agreement with
the degree of evolution discussed in \S~\ref{sec|results} found by
evolving the \mbh-\sis\ relation. Although these results are in
reasonable agreement with other works (Marconi et al. 2004; Merloni
et al. 2005; De Zotti et al. 2006; SWM; Merloni \& Heinz 2008),
uncertainties on the lower limit of the stellar mass function and/or
on the true fraction of stellar mass associated to BH growth at any
time, make this method less reliable than the one based on velocity
dispersion, and we therefore do not pursue it further.

\subsection{THE IMPACT OF MERGERS}
\label{subsec|mergers}

So far we have neglected mergers in our calculations.  Major mergers
between massive galaxies do occur, although recent work has
suggested the galaxy merger rate may be lower than previously
thought. Drory \& Alvarez (2008) compared the time variation in the
stellar mass function with the evolution implied by the star
formation rate alone, concluding that galaxies with stellar masses
above $10^{11}\, M_{\odot}$ undergo at most one major merger since
$z\sim 1.5$, in agreement with the results of Bell et al. (2007).
Lotz et al. (2006) find evidence for an even lower merger rate since
$z\sim 1$ from the DEEP2 survey.  Most importantly, however, a
significant rate of major mergers would strengthen our conclusions.
In velocity dispersion space, collisionless major mergers do not
significantly affect the final $\sigma$. For example, in a dry
merger of comparable--mass galaxies with mass $M_1$ and $M_2$ and
corresponding velocity dispersions $\sigma_1$ and $\sigma_2$, the
resulting galaxy will have a velocity dispersion $\sigma^2\sim
[M_1\sigma_1^2+M_2\sigma_2^2]/(M_1+M_2)\lesssim {\rm
  max}(\sigma_1^2,\sigma_2^2)$ (e.g., Ciotti et al. 2007). Therefore,
if the masses of the two galaxies are comparable, the final $\sigma$
will be close to the velocity dispersion of the progenitors.

\begin{figure*}[t!]
\epsscale{1.} \centering
\plotone{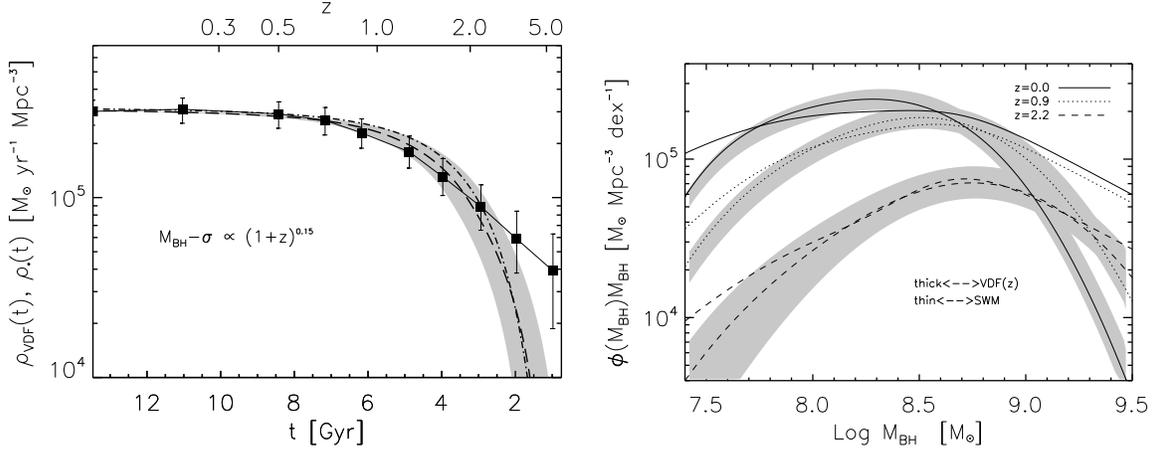} \caption{\emph{Left} panel: comparison between
the \rhoz\ and the \rhovdf\ computed for a model in which the
$\Phi(\sigma,z)$ is convolved with a \mbh-\sis\
  relation the scatter of which increases with redshift from $\eta=0.22$ to 0.40 . \emph{Right} panel:
comparison between the mass functions predicted from the same model
and from the AGN luminosity functions with $\lambda=0.6$.}
 \label{fig|ScatterRhoBH}
\end{figure*}

Dry mergers would then double the number of galaxies we predict at
fixed $\sigma$. Every dry major merger would in fact ``split'' the
galaxy into two (going back in time), adding an extra galaxy with
the same $\sigma$, compared to our present predictions (which
neglect mergers). In turn, this implies that the merger--free VDF
$\Phi(\sigma,z)$ computed above for $z>0$ is a \emph{lower} limit to
the true abundance of galaxies with velocity dispersion $\sigma$ at
redshift $z$. The associated BH mass density $\rho_{\bullet}(z)$
will consequently also be underestimated at redshift $z$. The
inclusion of any
mergers then predicts a larger BH mass density at fixed $\sigma$; to
compensate for this increase, a lower normalization of the
\mbh-\sis\ relation at $z=1-2$ is then required, which strengthens
our conclusions that large $\alpha$ values are excluded by the match
between \rhovdf$\,$ and \rhoz$\,$ (see the left panel in
Figure~\ref{fig|rhoBHz}).

However, if dissipation played a non--negligible role during the
evolution of the galaxy (either as a result of mergers, or in isolation),
then the velocity dispersion may increase with time
from the epoch of first collapse. To mimic such effects, we allow
all velocity dispersions to increase at higher redshifts as
$\sigma(z)=\sigma(0)\times(1+z)^{-\gamma}$. Most probably this
evolution is mass and/or velocity-dependent, nevertheless this
approach will be able to set interesting constraints on the mean
variation of $\sigma$. Also, any estimate for $\gamma$ should here
be considered as a lower limit to the actual evolution of $\sigma$,
as we neglect the still poorly understood increase in galaxy number
density due to possible galaxy mergers. In Figure~\ref{fig|wet} we
show the main results for a model with $\gamma=0.23$ and
$\alpha=1.5$.  Note that with the adopted scaling $M_{\rm BH}\propto
\sigma^\beta$ with $\beta\approx 4$, we expect, a degeneracy between
$\gamma$ and $\alpha$ given approximately by $\gamma \approx
\alpha/\beta \approx \alpha/4$ (although the degeneracy is modified
slightly by the assumed scatter and age--spread of BHs at a given
$z$ and $\sigma$).
We find that the downsizing evolution in this case is canceled out
(upper left panel), as all galaxies are now pushed to lower and
lower $\sigma$ at higher redshifts.

\begin{figure*}[t!]
\epsscale{1.} \centering
\plotone{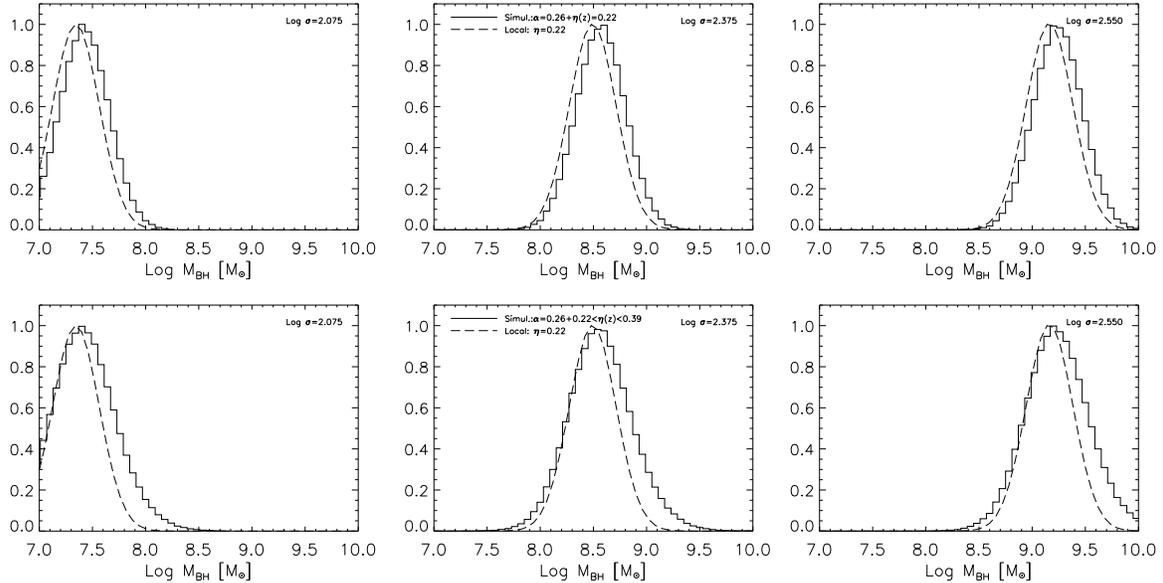} \caption{In each panel the solid histograms show
the mean of a set of Monte Carlo simulations which compute from our
models the expected distribution of black hole masses at fixed
velocity dispersion, as labeled (see text for details). The
long-dashed lines refer to the Gaussian distribution with
$\eta=0.22$ measured by Tundo et al. (2007). Both our models with
constant (upper panels) or evolving (lower panels) scatter still
produce at $z=0$ black hole mass distributions at fixed velocity
dispersion comparable to what is observed. Models characterized by
stronger redshift evolution will then evidently predict a scatter in
the local \mbh-\sis\ relation much larger than what is observed.}
\label{fig|ScatterBins}
\end{figure*}

The evolving number density in this model, shown in the upper right
panel of Figure~\ref{fig|wet}, seems to be at variance with the
number density evolution of early-type galaxies within $0\lesssim z
\lesssim 1$ inferred from DEEP2 by Faber et al. (2007; solid points
in the same Figure). However, lacking a clear understanding of how
the $\Phi(\sigma,z)$ should evolve in the presence of mergers, this
model cannot be ruled out, although some inconsistencies can already
be pointed out. From the Figure it can be seen that the number
density of the massive red galaxies in DEEP2 at the intermediated
redshifts of $0.5<z<1$, should be matched with galaxies
characterized by a velocity dispersion $\sigma\lesssim 100\, {\rm
km\, s^{-1}}$. Alternatively, the DEEP2 number densities could be
matched with the number density of galaxies with larger velocity
dispersions if mergers were a significant component in the evolution
of these galaxies, thus significantly increasing their number
density at higher redshifts. However, the latter hypothesis may
contradict independent results (e.g., Lotz et al. 2006). The
dissipative model described here also predicts a mean velocity
dispersion about flat out to $z\sim 2$ and slightly decreasing at
higher redshifts, as shown in the lower left panel of
Figure~\ref{fig|wet}.

The main achievement of this model is the good match between
\rhovdf\ and \rhoz\ even if a strong evolution in the \mbh-\sis\
relation has been assumed ($\alpha=1.5$), as shown in the lower
right panel of Figure~\ref{fig|wet}. This model is characterized by
a significant dissipative phase in the evolution of typical
early-type galaxies, which could represent an interesting constraint
for galaxy evolution models and it can in principle be tested
through hydrodynamical simulations, which we plan to do in future
work.

On the other hand, a major problem with the dissipative model is
represented by its implied Eddington ratio distribution. We in fact
find that the strong increase in the \mbh-\sis\ normalization at
higher redshifts requires a significant \emph{decrease}, by up to a
factor of a few, in the mean Eddington ratio $\bar{\lambda}(z)$ to
keep the match between the BH mass functions at $z\gtrsim 2$ shown
in Figure~\ref{fig|MFz}. The latter behavior of $\bar{\lambda}(z)$
is at variance with several works which actually claim an almost
constant or probably increasing $\bar{\lambda}(z)$ at higher
redshifts (e.g., McLure \& Dunlop 2004; Shankar et al. 2004;
Vestergaard 2004; Kollmeier et al. 2006; Netzer \& Trakhtenbrot
2007; SWM; Shen et al. 2008b; Shankar et al. 2008b).

\subsection{COMPARISON WITH PREVIOUS WORKS}
\label{subsec|previousworks}

The relatively mild \mbh-\sis\ redshift evolution inferred from our
approach may seem in apparent disagreement with some recent
independent studies. As briefly mentioned in \S~\ref{sec|intro},
Treu et al. (2007) and Woo et al. (2008) have randomly compiled from
the SDSS Data Release 4 a sample of about 20 Seyferts galaxies in
the redshift range $0.37\lesssim z \lesssim 0.57$. Their results,
shown as open circles in Figure~\ref{fig|MbhSigmaz}, are compared
with those of Shen et al. (2008, shown as filled circles), who
estimated the \mbh-\sis\ relation for a larger sample of active
galaxies up to $z=0.452$. While the latter claim that no significant
evolution in the \mbh-\sis\ relation is detectable from their
sample, Woo et al. (2008) confirm the results by Treu et al. (2007)
that a significant increase of $\sim 0.2$ dex in BH mass at fixed
velocity dispersion must occur within $z=0$ and $z\sim 0.5$. Our
best-fit model, shown at redshifts $z=0$ and $z=0.5$ with
long-dashed and solid lines respectively, shows no strong evolution
within this redshift range and it is in reasonable agreement with
both samples. A significant discrepancy is noticeable with respect
to the Woo et al. (2008) results for velocity dispersions $\log$
\sis$/$\kms$\lesssim 2.3$. However, systematic uncertainties may
affect these estimates; for example as also discussed by Woo et al.
(2008), especially in galaxies with lower BH mass, the host galaxy
contribution to the 5100{\AA} luminosity may lead to an
overestimation of the true BH mass. Overall, given the systematics
and biases which affect these kind of studies (e.g., Lauer et al.
2007), we do not find strong evidence for a disagreement between
these works and our results. For the same reasons, we do not attempt
comparisons with the results obtained from higher redshift studies.

\begin{figure*}[t!]
\epsscale{1.} \centering
\plotone{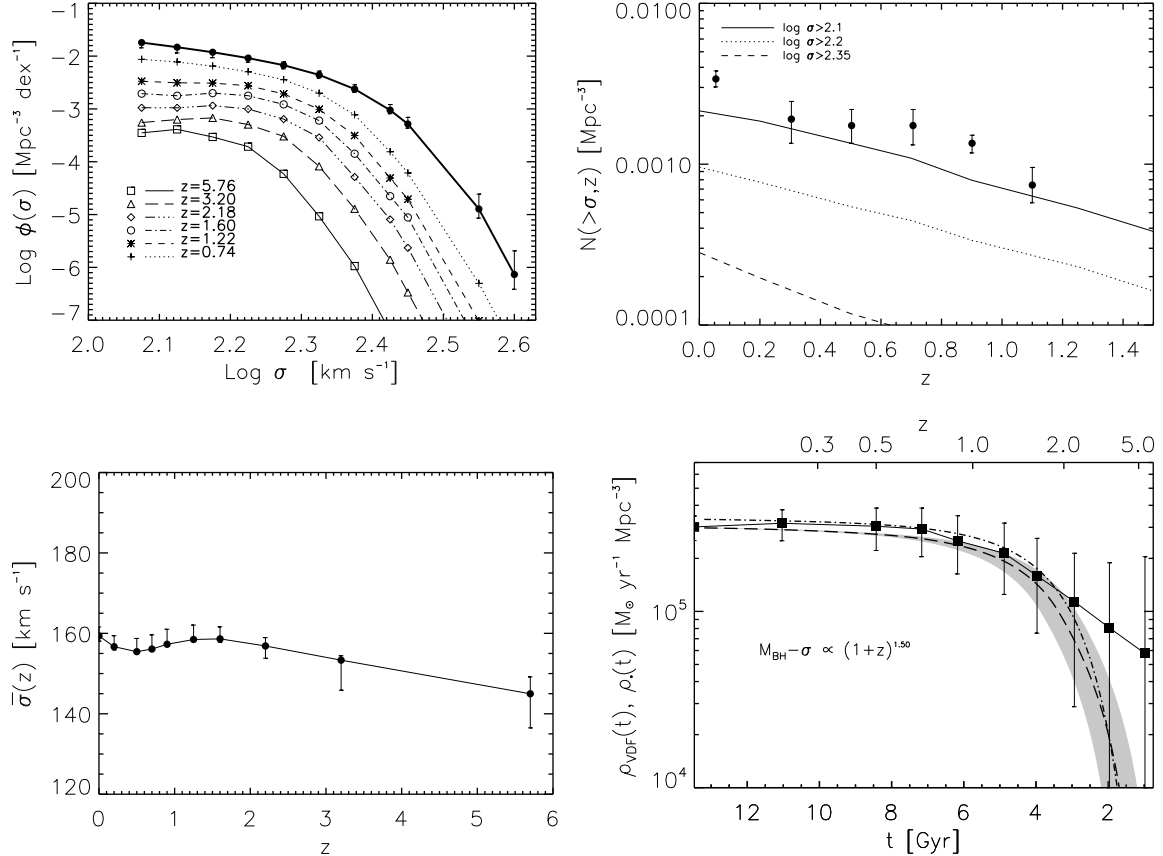} \caption{Results for a model in which we allow
the mean galaxy velocity dispersion to decrease at higher redshifts
to mimic the effects of prolonged wet activity in the host galaxies
since their formation epoch. We assume an evolution of the type
$\sigma(z)=\sigma(0)\times(1+z)^{-\gamma}$ with $\gamma=0.23$. The
downsizing effect is canceled out (\emph{upper left}) and the match
with DEEP2 number density evolution is significantly worsen
(\emph{upper right}). At variance with our previous results this
model predicts a mean velocity dispersion about flat out to $z\sim
2$ (\emph{lower left}) and, more important, a good match between
\rhovdf\ and \rhoz\ with $\alpha=1.5$ (\emph{lower right}).}
\label{fig|wet}
\end{figure*}

Merloni et al. (2004) compared the accreted BH mass density in AGNs
with the cosmological global star formation rate density (see also
Haiman et al. 2004). Although their conclusions depend on additional
assumptions about the fraction of the star forming galaxies which
are linked to BH growth at a given redshift, irrespective of the
adopted value of the radiative efficiency their best-fit relation
yields $\alpha\approx 0.5$, somewhat higher, but still consistent,
with the value found here, and they also rule out $\alpha\gtrsim
1.2$ at a high confidence level. Hopkins et al. (2006) also describe
a model-independent integral constraint that defines an upper limit
to the allowed degree of evolution in the ratio of BH mass to host
galaxy luminosity or mass, as a function of redshift. By comparing
the AGN density with the luminosity and mass functions in different
bands from redshifts $z=0-2$, they rule out at $\gtrsim 6\sigma$ a
BH-host galaxy mass ratio significantly larger at high redshifts
than locally. Cattaneo \& Bernardi (2003) combined a relation
between mean age and velocity dispersion, derived from a sample of
SDSS local early-type galaxies, with the Sheth et al. (2003) local
VDF. By assuming a redshift independent mean Eddington ratio,
radiative efficiency and obscuration correction, they were then able
to reproduce the AGN optical and X-ray luminosity functions. As
mentioned above, similar calculations have been performed recently
by Haiman et al. (2007), whose results imply, assuming a
non--evolving \mbh-\sis\ relation, that in order to reproduce the
bolometric quasar luminosity function, the quasars must shine at a
mean sub-Eddington regime of $\lambda=0.5$ that is approximately
constant with time.  This conclusion was confirmed by the
independent estimates of SWM. However, the works by Cattaneo \&
Bernardi (2003) and Haiman et al. (2007) can only constrain the {\it
combination} of the Eddington ratio distribution and the \mbh-\sis\
relation, while our approach here can simultaneously constrain the
mean accretion histories of BHs and their host galaxies, and the
mean Eddington ratio of BHs at all times. A further difference is
that the analysis of Haiman et al. (2007) can constrain the quasar
lifetime, while the results here rely on the comparison between
time--integrated quantities, and are strictly independent of the
quasar lifetime.

\begin{figure}[t!]
\epsscale{1.} \centering
\plotone{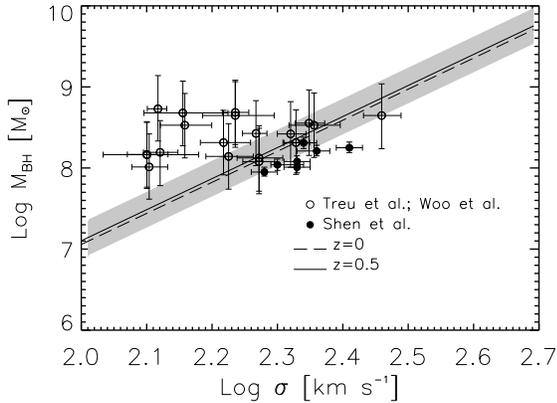} \caption{Our best-fit model for the \mbh-\sis\
relation is plotted at $z=0$ and $z=0.5$, as labeled, and compared
with recent data from Treu et al. (2007) and Woo et al. (2008),
shown with \emph{open} symbols, and Shen et al. (2008), shown with
\emph{filled} symbols.} \label{fig|MbhSigmaz}
\end{figure}

\section{CONCLUSIONS}
\label{sec|conclu}

In this work we combined the local VDF with the stellar age
distributions estimated by Bernardi et al. (2006), to compute the
VDF at higher redshifts, $\Phi(\sigma,z)$.  In agreement with
previous work, we find statistical evidence for downsizing, whereby
the stellar populations in galaxies with higher velocity dispersions
formed earlier, irrespective of the specific model we adopt for
computing the galactic ages. We then computed the BH mass function
associated with $\Phi(\sigma,z)$ at each redshift $z$, through a BH
mass -- velocity dispersion relation whose normalization was allowed
to evolve with redshift as $\propto (1+z)^{\alpha}$.  Our main
underlying assumptions are that most of the growth of the central BH
occurs simultaneously (within $\pm$ 1 Gyr) with the formation of the
host's potential well, and that the measured stellar ages represent
this formation time to within a similar accuracy.  The BH mass
density \rhovdf$\,$ inferred from the VDF can then be compared with
the accumulated BH mass density implied by the time--integral of the
AGN luminosity function, \rhoz$\,$.  We find significant evidence
that the match between \rhovdf$\,$ and \rhoz$\,$ implies a
relatively mild redshift evolution, with $\alpha\lesssim 0.30$, and
with values of $\alpha\gtrsim 1.3$ excluded at 99\% confidence. If a
positive redshift evolution stronger than $\alpha\gtrsim 1$ were to
be confirmed independently in the future, then this would be a
robust indication that dissipative processes played a significant
role in galaxy evolution, resulting in an increase in the velocity
dispersion of the spheroid components of individual galaxies with
cosmic time. However, we also find evidence that a dissipative model
predicts a mean Eddington ratio decreasing with increasing redshift,
at variance with several independent studies.

\acknowledgements This work was supported by NASA grants
GRT000001640 (to FS), NNG04GI88G and NNX08AH35G (to ZH) and LTSA-NNG06GC19G (to
MB). ZH also acknowledges support by the Pol\'anyi Program of the
Hungarian National Office of Technology. FS thanks David H. Weinberg
for interesting discussions.


{}

%


\end{document}